\documentclass[%
 reprint,
 superscriptaddress,
 amsmath,amssymb,
 twocolumn,
 aps,
 pra,
]{revtex4-2}

\usepackage{graphicx}
\usepackage{dcolumn}
\usepackage{bm}
\usepackage{siunitx}

\begin{document}

\preprint{APS/123-QED}

\title{Digital synchronization for continuous-variable quantum key distribution}

\author{Hou-Man Chin}
 \altaffiliation[Also at ]{Machine Learning in Photonic Systems, Department of Photonics, Technical University of Denmark, 2800 Lyngby, Denmark}
 \email{homch@dtu.dk}
\author{Nitin Jain}
\affiliation{Center for Macroscopic Quantum States (bigQ), Department of Physics, Technical University of Denmark, 2800 Lyngby, Denmark}
\author{Ulrik L. Andersen}
\affiliation{Center for Macroscopic Quantum States (bigQ), Department of Physics, Technical University of Denmark, 2800 Lyngby, Denmark}
\author{Darko Zibar}
\affiliation{Machine Learning in Photonic Systems, Department of Photonics, Technical University of Denmark, 2800 Lyngby, Denmark}
\author{Tobias Gehring}
\email{tobias.gehring@fysik.dtu.dk}
\affiliation{Center for Macroscopic Quantum States (bigQ), Department of Physics, Technical University of Denmark, 2800 Lyngby, Denmark}%

\date{\today}

\begin{abstract}
Continuous variable quantum key distribution (CV-QKD) is a promising emerging technology for the distribution of secure keys for symmetric encryption. It can be readily implemented using commercial off-the-shelf optical telecommunications components. A key requirement of the CV-QKD receiver is the ability to measure the quantum states at the correct time instance and rate using the correct orthogonal observables, referred to as synchronization. We propose a digital synchronization procedure for a modern CV-QKD system with locally generated local oscillator for coherent reception. Our proposed method is modulation format independent allowing it to be used in a variety of CV-QKD systems. We experimentally investigate its performance with a Gaussian modulated CV-QKD system operating over a 10-20 km span of standard single mode fibre. Since the procedure does not require hardware modifications it paves the way for cost-effective QKD solutions that can adapt rapidly to changing environmental conditions.
\end{abstract}

\maketitle


\section{\label{sec:intro}Introduction}

Quantum key distribution (QKD) is a future proof information theoretically secure method of generating and distributing cryptographic keys over a communication link, assumed to be fully under the control of an eavesdropper (Eve)~\cite{Diamanti2015, Pirandola2020}. First proposed in \cite{Bennett1984}, it relies on fundamental laws of physics to enable data encryption with information theoretic security. Quantum information is often encoded onto quadrature amplitudes of light which are then detected by a coherent receiver to yield continuously varying outcomes. QKD based on those quadrature amplitudes is known as continuous-variable (CV) QKD~\cite{Ralph1999}, and has been successfully demonstrated both in laboratory conditions~\cite{Grosshans2003, Lance2005, Jouguet2013, Eriksson2019, Chin2021} and field trials~\cite{Huang2016, Zhang2019, Aguado2019}.

\begin{figure}
    \centering
    \includegraphics[width=0.48\textwidth]{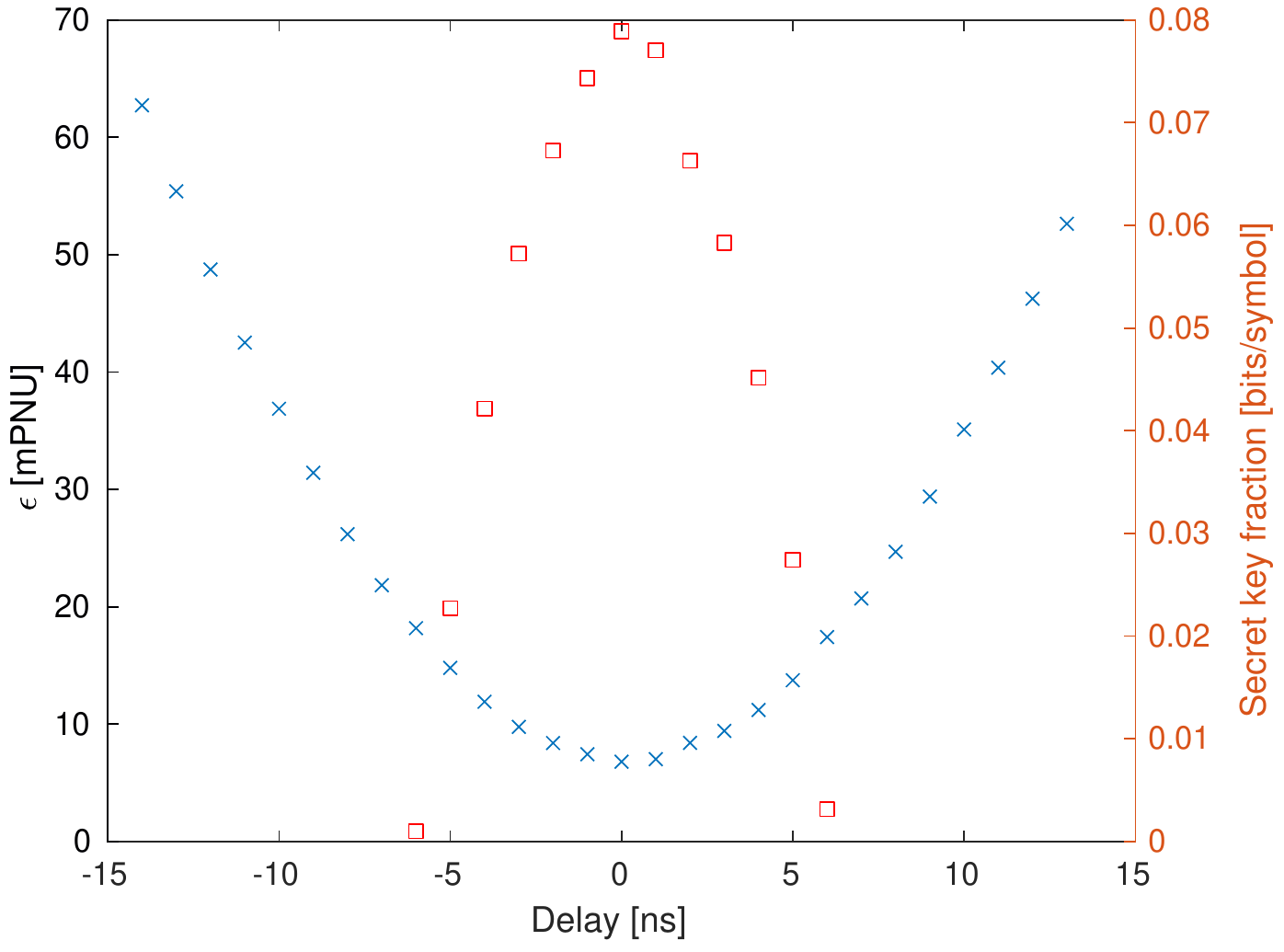}
    \caption{Impact of forced erroneous synchronization on excess noise and secret key fraction in a clock synchronous system transmitting at 20 MSymbols per second over 20km fibre.}
    \label{fig:synceddelay}
\end{figure}

Any CV quantum communication system must contend with synchronization. In essence, this involves ensuring that the recipient (Bob) is looking at what the sender (Alice) is transmitting such that in the ideal scenario Bob's information only differs from Alice's by the fundamental quantum noise. Any imperfections in the synchronization process will however create additional noise, so-called \textit{excess noise}, which must be attributed to Eve. Since excess noise reduces the secret key length until key generation eventually fails, it is desirable to reduce noise from imperfect synchronization as much as possible. 
Figure~\ref{fig:synceddelay} shows an example of the effect of forced incorrect time delay synchronization on excess noise and respective secret key fraction in an experimental CV-QKD measurement. Here, a compensation of the synchronization delay that is off the true value by more than about 1/10 of the symbol duration prevents secret key generation. This example gives an idea of the tight margin in which the QKD system has to operate in.

Traditionally, CV-QKD systems transmit the local oscillator (LO) through the same fiber as the encoded quantum states by using time and polarization multiplexed optical pulses. Such systems tap off a small fraction of the (powerful) LO for clock synchronization at the receiver~\cite{Jouguet2013,Zhang2019,Zhang2020}. With the LO generated by a separate laser at the receiver, this is no longer an option. Instead, a convenient method for Alice and Bob to perform clock synchronization is to transmit and detect, together with the quantum information at 1550\,nm, a clock tone on a significantly different wavelength, such as 1300\,nm \cite{Eriksson2018}. The disadvantage of this method is of course requiring another transmitter (Tx) and receiver (Rx) at this wavelength. Many proof-of-principle experiments in the lab therefore resort to electrical synchronization of the waveform generator and the data acquisition device~\cite{Chin2021, Milovancev2021, Roumestan2021, Laudenbach2018, Jain2021}. 

Here, we propose to perform digital synchronization with the aid of classical pilot tones and a quadrature phase shift keying (QPSK) signal. These are frequency multiplexed to the coherent states of the quantum signal and can be generated by the same waveform generator. Thus, our method does not require additional hardware. It is furthermore designed such that it does not require pre-calibration~\cite{Kleis2017} and is independent of the modulation format~\cite{Kleis2019b} and modulation variance, i.e.\ it is suitable for low-order quadrature amplitude modulation~\cite{Denys2021}, Gaussian modulation~\cite{Pirandola2021} and also for measurement-device independent systems~\cite{Pirandola2015}. Such pilot tones are already often used for phase noise compensation \cite{Kleis2017, Laudenbach2019, Chin2021}.

We implement such a procedure experimentally and examine its performance using a CV-QKD setup. We show that it allows a realistic free-running configuration to achieve similar performance to an externally synchronized one, over both a 10\,km and a 20\,km optical fibre as the quantum channel connecting Alice and Bob.

\subsection{\label{sec:clkRecovery}Synchronization}

Synchronization is a quintessential component of any communication system. In an optical quantum communication system, we distinguish between three effects that must be compensated for, to achieve synchronization.
First, is of course the signal propagation through the channel, which produces a delay between the transmission and reception. Second, is the difference in signal period, i.e.\ what each party defines as one hertz and last is the drift in the optical phase, typically dominated by laser phase noise.

In telecommunications, before the advent of the digital coherent receiver, signal synchronization was performed in hardware with devices such as phase locked loops which corrected the timing, phase and frequency offset between the transmitter and receiver clocks. A timing error signal is generated from the sampled signal which then feeds into a voltage controlled oscillator (VCO) sampling the received signal. The introduction of the digital coherent receiver meant that it was possible to perform clock recovery in the digital domain, allowing for digital generation of the timing error signal and replacing the VCO with digital interpolation. 

Typical digital telecommunication synchronization schemes first compensate for the relative difference in sampling period, then frequency and phase error, and lastly for the channel delay. The first is fixed by timing error detectors \cite{Gardner1986, Mueller1976} which estimate how incorrect a sampling point is, and generate an error signal. From this, the receiver can converge towards an optimum sampling point. This process is called \textit{re-timing} or \textit{clock recovery}. Frequency offset compensation requires simply to find the highest power within the signal bandwidth and then frequency downshifting it to baseband. Next, phase noise compensation is performed through estimation of the phase noise using digital signal processing (DSP) such as the $M^{th}$ power algorithm \cite{Faruk2017} or pilot tones \cite{Chin2021}. The channel delay then becomes a matter of finding the start of the transmitted signal which can be done with cross correlation of a known sequence.

These methods are (see \cite{Mueller1976, Gardner1986, Faruk2017}) usually performed on the received signal itself with minimum additional overhead in frequency or time. Performing digital synchronization for a CV-QKD system poses unique challenges to these schemes. Firstly, the above mentioned methods require a discrete modulation format with well defined temporal features, however the CV-QKD system may not be using such a discrete modulation format as required by the implemented security proof \cite{Pirandola2021}. Secondly, the typical signal to noise ratio (SNR) of a quantum signal at the receiver is less than 0 dB \cite{Jouguet2011, Huang2016b} restricting the accuracy of those methods. These are fundamental limitations, negating the possibility of using a timing error detector to determine the optimum sampling point.
To solve these challenges we resort to auxiliary signals with higher power as we will see in the following.

\section{\label{sec:clkAlgo}Digital synchronization algorithm}

\begin{figure}[htb]
    \centering
    \includegraphics[width=0.5\textwidth]{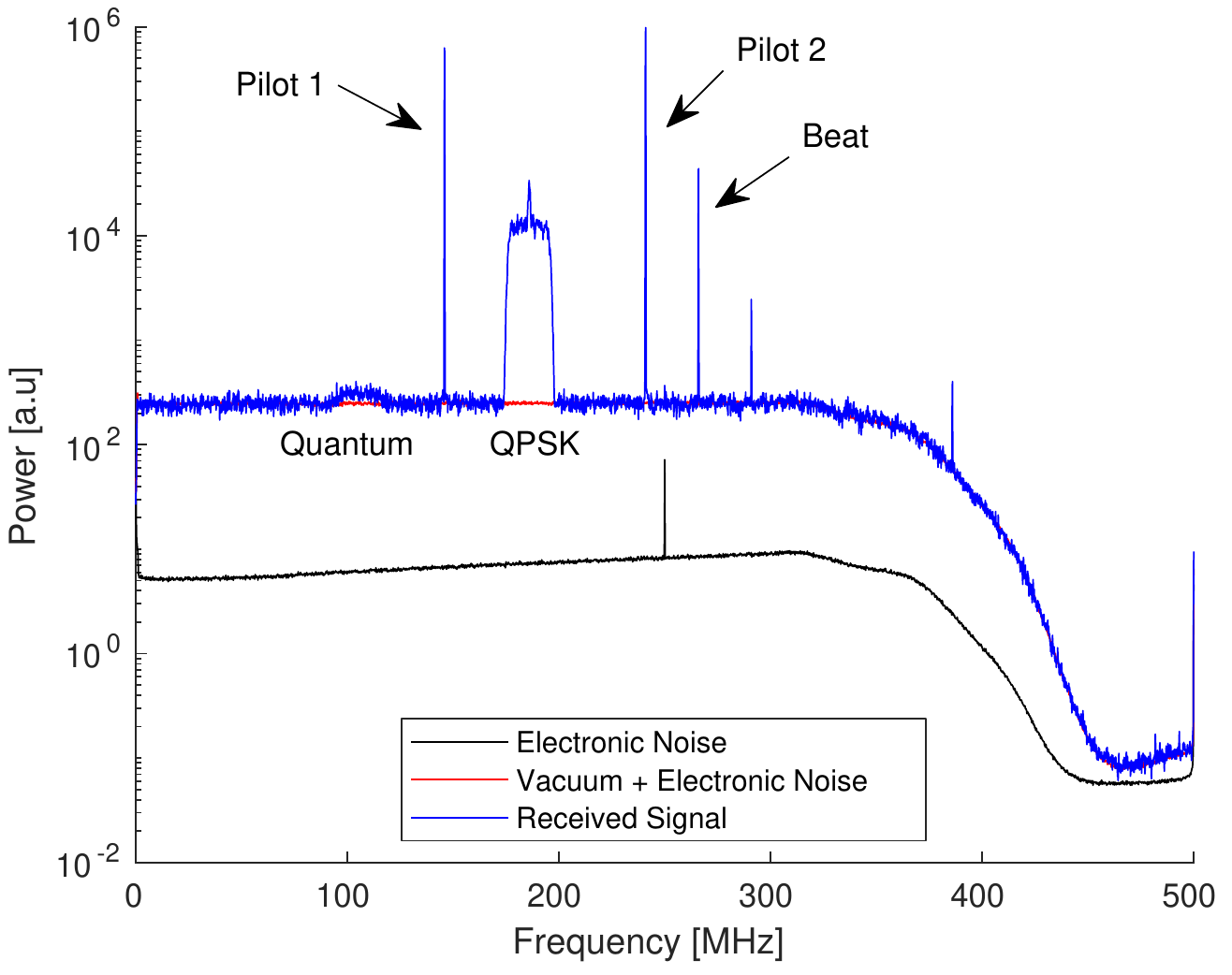}
    \caption{Various spectra obtained from the CV-QKD system operating over a 20km long quantum channel. The center of the quantum signal lies at approximately 105 MHz, and the center of the QPSK signal at 185 MHz.}
    \label{fig:psd}
\end{figure}

\begin{figure}[htb]
    \centering
    \includegraphics[width=0.3\textwidth]{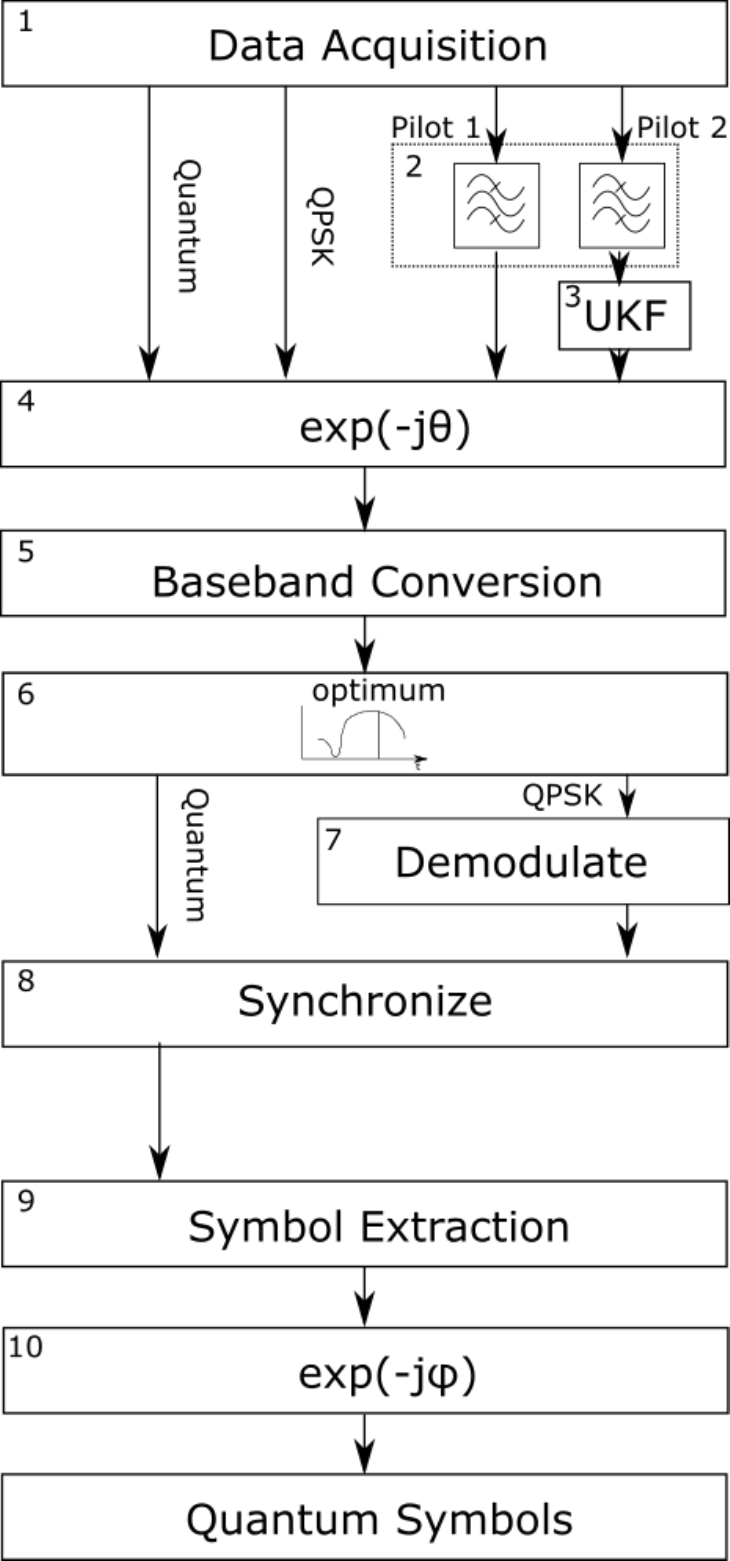}
    \caption{DSP chain for our CV-QKD receiver.}
    \label{fig:clksyncblocks}
\end{figure}

\begin{figure*}[htb]
    \centering
    \includegraphics{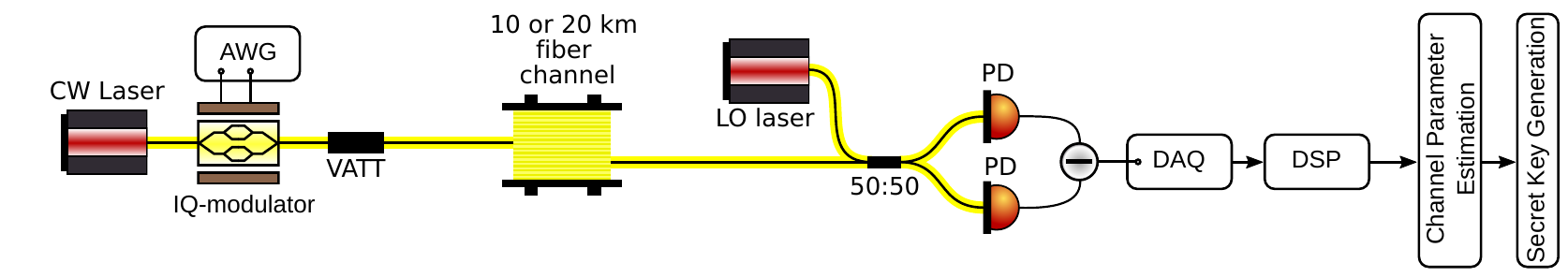}
    \caption{Experimental setup modulating Gaussian coherent states using an in-phase and quadrature (IQ) electro-optic modulator onto a continuous wave (CW) laser. A variable optical attenuator (VATT) reduces the mean photon number of the coherent state ensemble in the quantum signal to the order of 1 photon. Transmitter and LO lasers are specified to be sub 100 Hz linewidth. A 1 GSample/s arbitrary waveform generator (AWG) and data acquisition (DAQ) perform respectively signal generation and digitization at the transmitter and receiver. Digital signal processing (DSP) performs digital synchronization and raw key generation as described in the main text. The processing is performed offline. Channel parameter estimation is performed on the data and estimates excess noise and channel transmission. Secret key generation distills the secret key from the raw key.}
    \label{fig:expsetup}
\end{figure*}

Before we discuss the DSP synchronization algorithm, let us describe the involved frequency multiplexed signals. Figure~\ref{fig:psd} shows an exemplary power spectrum of the sampled output of the balanced receiver used for heterodyne detection with a local oscillator whose frequency is shifted with respect to the transmitter laser by about 265 MHz. All involved signals are generated as optical single sidebands of the transmitter laser. 

The quantum signal, located at roughly 105 MHz in the modulated spectrum in Fig.~\ref{fig:psd}, contains root-raised-cosine pulse shaped coherent states with random quadrature amplitudes sampled from, in our case, a complex Gaussian distribution. As the proposed synchronization method works independently of the chosen modulation format, it could however be from any other distribution. The outcomes of in-phase and quadrature (amplitude and phase quadrature) measurements of the coherent states at the receiver constitute the raw key which is distilled into the secret key by the data processing phase of the QKD protocol. The power of the quantum signal must be optimized with respect to the channel loss and QKD protocol parameters to maximize the secret key yield and has a typical SNR below 0 dB at the receiver.

For digital synchronization we temporally multiplex some coherent states with known complex amplitude as reference symbols using a constant-amplitude-zero-autocorrelation (CAZAC) sequence. Furthermore, in the frequency domain, the quantum signal accompanies two pilot tones and a quadrature phase shift keying (QPSK) signal of higher power than the quantum signal. The QPSK signal contains a sequence of known symbols as well as an ID number for the frame.

Equipped with these four signals, the digital signal synchronization procedure performed at the receiver is as shown in Fig.~\ref{fig:clksyncblocks}. If the transmitter and receiver share the same clock, the pilot signal position relative to the quantum signal is known exactly, thus $\text{freq}_\text{offset}^\text{tx} = \text{freq}_\text{offset}^\text{rx}$. When the clocks are free-running however, the immediate effect is that the frequency location of the quantum signal with respect to the pilot at the receiver is not exactly as modulated at the transmitter, i.e.\ $\text{freq}_\text{offset}^\text{tx} \approx \text{freq}_\text{offset}^\text{rx}$, since Alice and Bob's \textit{truth} of what is one hertz is different. The requirement is to align Bob's measured signal in frequency and phase. To do this, we must first estimate the change in frequency offset per hertz caused by the free-running nature of the system. For this, we use both pilot tones to estimate 
\begin{equation}
    \Delta f = \frac{f_\text{pilot}^{1} - f_\text{pilot}^{2}}{\Delta f_\text{pilots}^{1,2}}\ ,
\end{equation}
where $f_\text{pilot}^{i}$ are estimated pilot tone frequencies at the receiver and $\Delta f_\text{pilots}^{1,2}$ is the frequency difference as modulated by the transmitter by filtering them in block 2 of Fig.~\ref{fig:clksyncblocks}. 
This modifier is then applied to the transmitted frequency offset $\text{freq}_\text{offset}^\text{tx}$ to calculate the received quantum and QPSK signal offsets from the pilot tone. We note that the beat between transmitter and receiver lasers may be used in lieu of a pilot tone, however we experienced occasional inconsistency and therefore opted to simply insert a second pilot tone.

Next, we compensate for optical phase noise from the transmitter and receiver lasers Fig.~\ref{fig:clksyncblocks}(3). The phase of the received signal is estimated using a machine learning framework implementing unscented Kalman filtering (UKF), from one of the pilot tones \cite{Chin2021} and then used to compensate all detected signals using an $\exp(-j\phi)$ operation Fig.~\ref{fig:clksyncblocks}(4).

Since there is a large degree of oversampling in our system as is typical for a heterodyne CV-QKD implementation, the optimum sample per measurement frame is calculated by picking the sampling point with maximum power in the QPSK signal after downconversion to baseband Fig.~\ref{fig:clksyncblocks}(5) and (6). This maximizes the mutual information of the QPSK signal, and therefore of the quantum signal since both signals are simultaneously generated at the transmitter.

The QPSK symbols are then extracted Fig.~\ref{fig:clksyncblocks}(7) after being match filtered i.e.\ the same anti-aliasing filter is used for the downsampling operation from the sampling rate of the receiver to symbol rate, as was used for signal upsampling at the transmitter from symbol to the sample rate of our arbitrary waveform generator. An $\text{M}^{th}$ power algorithm \cite{Leven2007} is applied to remove residual phase errors,
\begin{equation}
    \theta_{k} = \frac{1}{M} \text{arg}\left(\sum_{n=0}^{N-1}(x_\text{in}(n))^{M}\right)\ , 
    \label{eqn:Mth}
\end{equation}
where M in this case is 4 due to QPSK and $x_{in}$ are the received symbols, $N$ is the window length over which the phase error is calculated and $n$ is the current symbol. A known QPSK sequence of symbols is cross correlated with the extracted symbols to acquire the required synchronization delay Fig.~\ref{fig:clksyncblocks}(8). The quantum signal is then synchronized in sample space using the combination of this delay (converted to samples) and the previously obtained optimum sample. 

We note that the $\text{M}^{th}$ power algorithm is used in optical telecommunications for frequency offset estimation as well, using the change in phase between symbols after removing the modulation \cite{Faruk2017}. We instead use pilot tones for ease of estimation since it can be done prior to demodulation.

Next, the quantum signal is frequency shifted to baseband and downsampled to symbols with matched filtering Fig.~\ref{fig:clksyncblocks}(9). The phase output of the $\text{M}^{th}$ power algorithm is applied to the symbols as well. Lastly, the CAZAC symbols in the quantum signal are used to correct for residual bulk phase rotation Fig.~\ref{fig:clksyncblocks}(10).

\subsection{Experimental Setup}
\begin{figure*}[htb]
    \centering
    \begin{minipage}[b]{0.6\columnwidth}
        \centering
        \includegraphics[width=\textwidth]{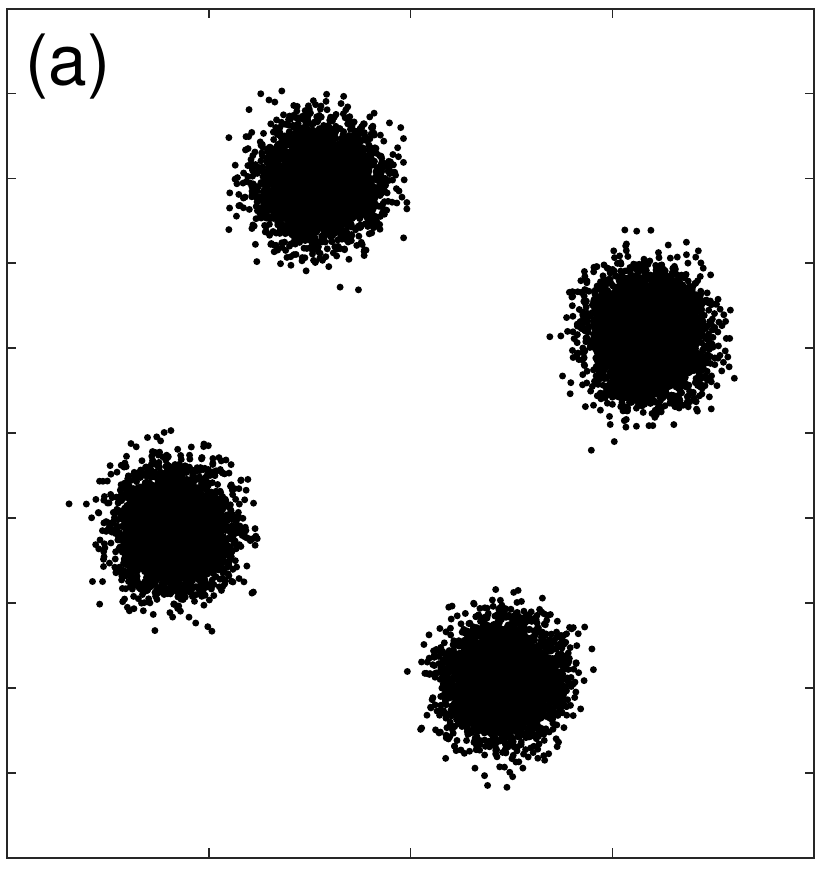}
    \end{minipage}
    ~
    \begin{minipage}[b]{0.6\columnwidth}
        \centering
        \includegraphics[width=\textwidth]{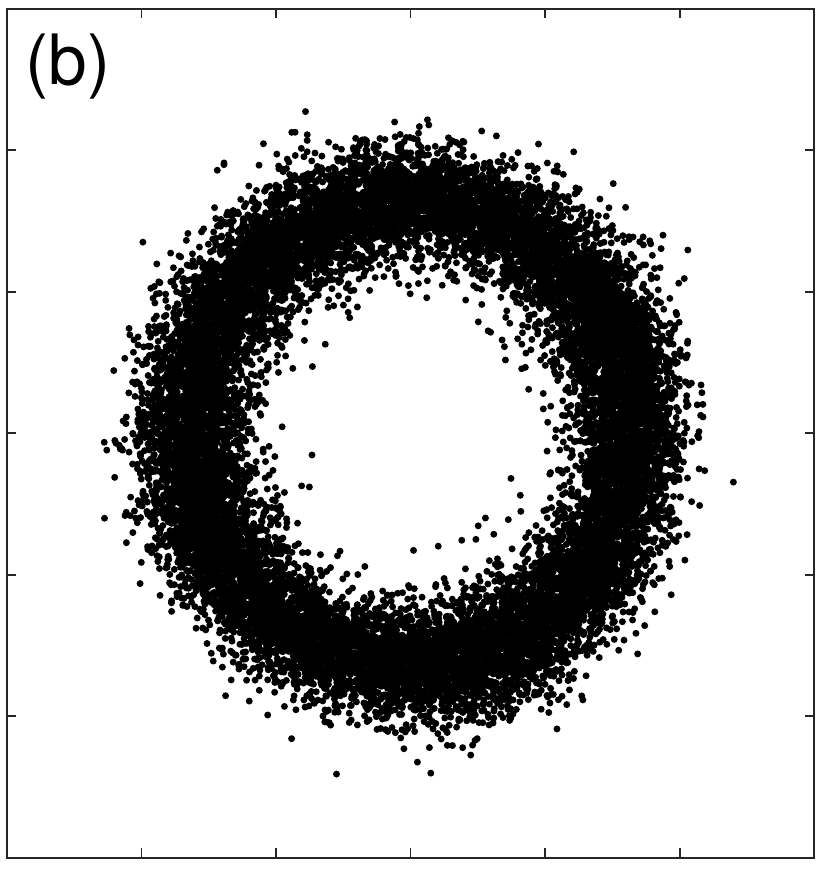}
    \end{minipage}
    \begin{minipage}[b]{0.6\columnwidth}
        \centering
        \includegraphics[width=\textwidth]{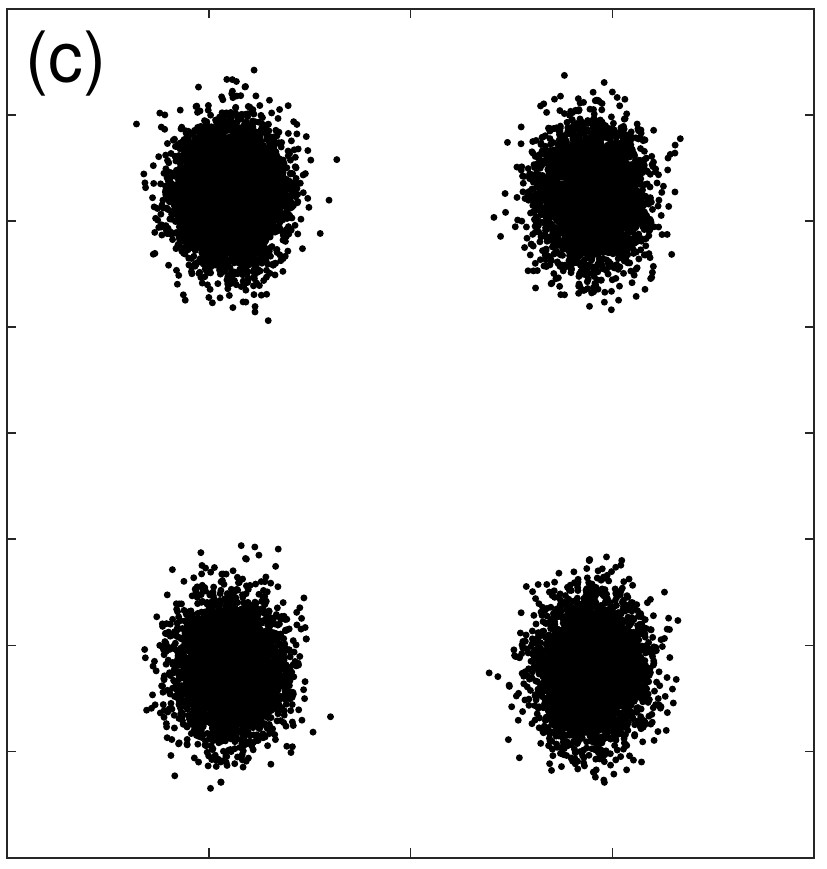}
    \end{minipage}
    \caption{After applying phase compensation using UKF, QPSK constellation with (a) shared reference clock, (b) internal (separate) clocks and (c) after digital synchronization procedure.}
    \label{fig:qpskconsts}
\end{figure*}
\begin{figure}[htb]
    \centering
    \includegraphics[width=0.45\textwidth]{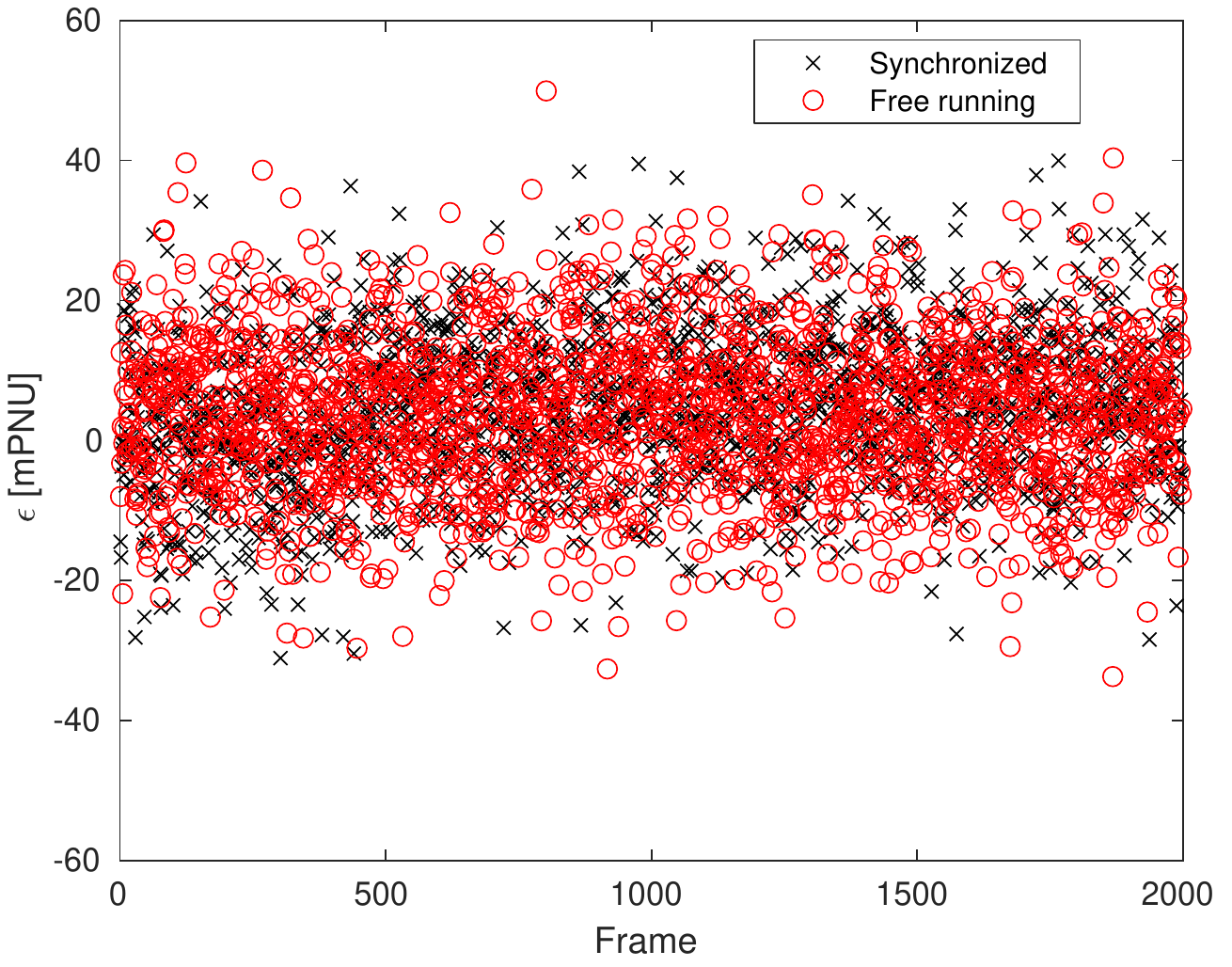}
    \caption{Excess noise (in mPNU) per frame for externally synchronized clock (which entails a shared reference clock between transmitter and receiver) and free-running systems.}
    \label{fig:mepperframe}
\end{figure}
\begin{figure}[htb]
    \centering
    \includegraphics[width=0.45\textwidth]{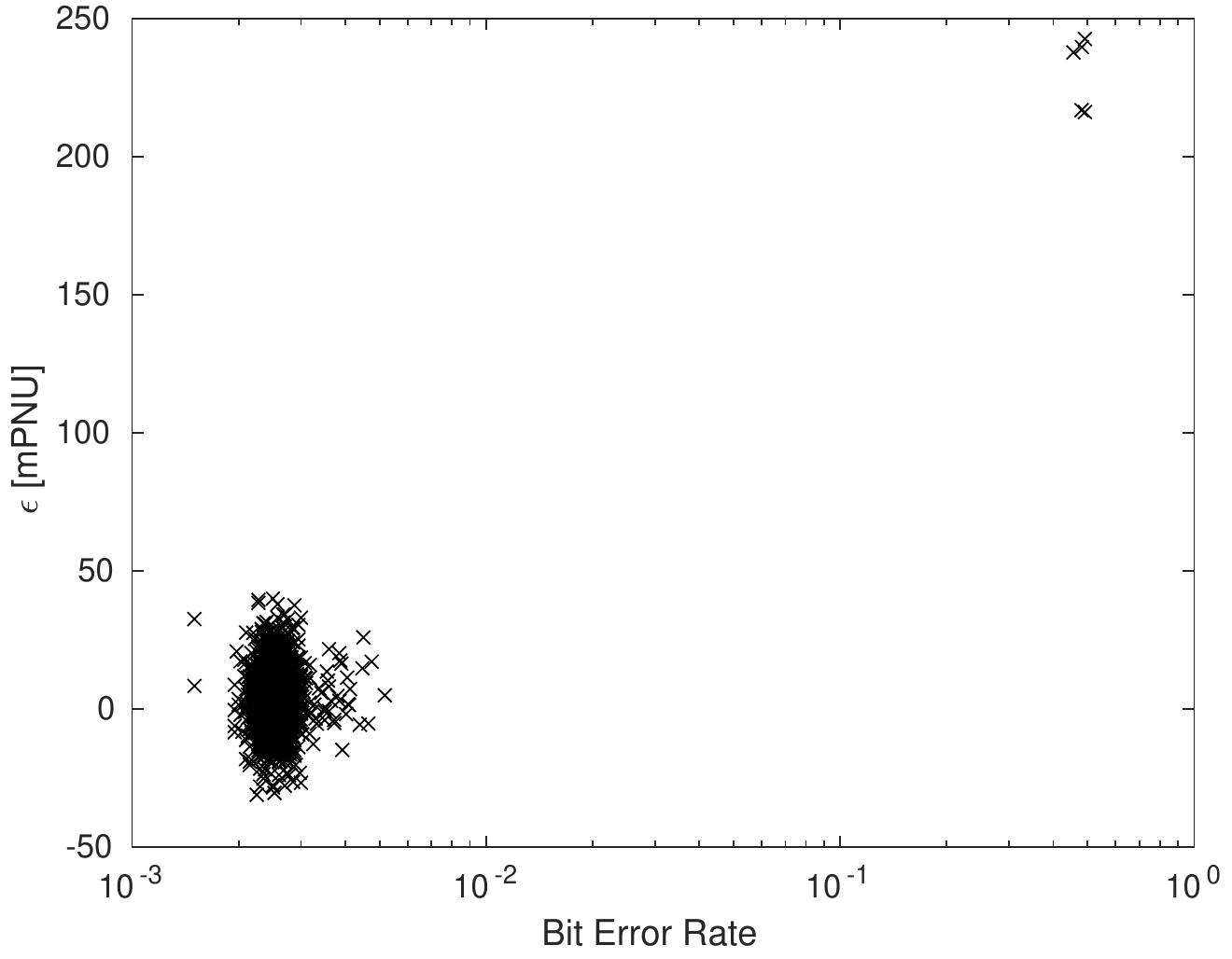}
    \caption{Excess noise measured in the quantum signal versus respective bit error rate of the QPSK signal for the frames transmitted and detected over a 20km long quantum channel.}
    \label{fig:bervsepn}
\end{figure}
To verify our digital synchronization scheme experimentally, we used a CV-QKD setup as shown in Fig.~\ref{fig:expsetup}. Initial testing was performed over 20\,km of standard single mode fibre (SMF-28) with the data acquisition (DAQ) in the receiver being externally triggered by the transmitter. Another measurement for confirmation was performed over 10\,km SMF, but this time without any external trigger; instead we used the DAQ channel itself as a trigger source \cite{Jain2022}. Alice modulates symbols drawn at random from a Gaussian distribution at 20 Mbaud. These digital symbols were upsampled to the 1 GSample/s sampling rate of the arbitrary waveform generator (AWG), with a root raised cosine anti-aliasing filter (roll-off = 0.2) after which they were frequency shifted by $f_\text{q} = 160$ MHz, i.e.\ multiplied with $\exp(j2\pi f_\text{q}t)$, for single sideband modulation. Two reference pilot tones were frequency multiplexed at $\tilde{f}_\text{pilot}^{1} = 120$ MHz and $\tilde{f}_\text{pilot}^{2} = 25$ MHz to the quantum signal. An additional 20 Mbaud QPSK signal was inserted at $\tilde{f}_\text{QPSK} = 80$ MHz. This radio frequency signal and a $\pi/2$-phase shifted version thereof drove the two arms of an in-phase and quadrature (IQ) electro-optical modulator to simultaneously modulate the signal in both quadratures onto the output of a laser centered at 1550.13 nm. The optical signal was then attenuated such that the mean photon number from only the quantum signal (i.e.\ excluding the pilot tones and QPSK signal) was $\approx$ 1.45 at the quantum channel input to the 20\,km and 3.5 for the 10\,km measurement.

At the channel output, the transmitted optical signal was detected using a balanced heterodyne coherent receiver with a free-running LO generated by a laser separate from the transmitter with an offset frequency $\approx$ 280\, MHz. The output of the balanced receiver was sampled at 1 GSample/s giving a spectrum similar to Fig.~\ref{fig:psd}.
The transceiver setup was synchronized with a 10 MHz clock reference for benchmarking purposes before being operated in a free-running configuration. The optical efficiency of the balanced receiver (due to the non-unity quantum efficiency of the photodiodes and optical loss from connectors) was measured to be $\approx 0.69$. The 10\,km measurement was performed using a similar system though the optical efficiency was instead $\approx 0.5$. 

The measurement time was divided into frames, each consisting of 10k Gaussian distributed complex values, or the `quantum symbols'. Transmitting in parallel, the first 2000 QPSK symbols were used for clock recovery and synchronisation in addition to 2000 reference symbols time multiplexed in the quantum signal. The QPSK symbols were transmitted at $\approx 7$ dB higher power than the quantum signal. The reference symbols that were time multiplexed with the quantum signal maintained the same power as the key symbols. Approximately 2000 data frames were detected. 
Additionally for the 10\,km measurement, the QPSK signal included frame ID data after the header sequence.

Initial calibration of the CV-QKD system was performed by measuring the electronic noise of the receiver by turning transmitter and receiver lasers off. The combined vacuum noise and electronic noise was then measured after turning on the LO, allowing for the normalization of the covariance matrix. Finally, the modulation variance was calibrated using the transmitter and receiver connected in a back to back configuration.

\subsection{Results}
First, to establish a baseline performance, we performed the exact same DSP process on a measurement set recorded with the QKD system which was synchronized using an external 10 MHz reference clock and an external trigger input to the receiver (first column in Table~\ref{tab:res}). The measurement was performed over 20 km fiber.
The auxiliary QPSK signal after phase noise compensation using the UKF had a constellation as shown in Fig.~\ref{fig:qpskconsts}a. The residual rotation of the constellation is due to a phase offset between the pilot tone used for phase estimation and the QPSK signal.
Next, we used a set of measurements from the free-running system (with the acquisition system being externally triggered). Using just the pilot tone for frequency offset estimation and phase compensation, as in Fig.~\ref{fig:qpskconsts}a, leads to the constellation in Fig.~\ref{fig:qpskconsts}b. Clearly, the QPSK constellation could not be recovered which is due to residual frequency offset error and clock run off. This is compensated by running the full DSP chain and the resulting constellation is shown in Fig.~\ref{fig:qpskconsts}c. The constellation is perfectly recovered and does also not show a residual phase rotation. 
\begin{center}
    \begin{table}
        \begin{tabular}{c|c|c|c}
         Measurement   &  1 & 2 & 3\\
         \hline
        Quantum Channel Length [km]         & 20      & 20      & 10\\
        Shared Clock Reference              & Yes       & No       & No\\
        Trigger Source, DAQ                & Ext.   & Ext.     & Chan.\\
        Modulation Strength [PNU]         & 1.45      & 1.45      & 3.50\\
        Excess Noise [mPNU]                 & 4.5       & 4.0         & 2.4\\ 
        Secret Key Fraction [bits/symbol]   & 0.0416    & 0.0466    & 0.0740
        \end{tabular}
    \caption{Summary of results from the three measurement sets. Ext.: Using external trigger input on the DAQ, Chan.: Using DAQ channel itself as the source of trigger.}
    \label{tab:res}
    \end{table}
\end{center}

As a performance metric we use the excess noise $\epsilon$, or the thermalization of the coherent states at the output of the channel. The thermalization is characterized by the mean photon number of the thermal state and specified in photon number units (PNU) in the following~\cite{Chin2021}. The results for the externally synchronized and the free-running systems are shown in Fig.~\ref{fig:mepperframe} over 20 km. Over the $\approx 1.9\times10^{7}$ symbols recovered, the average (channel output related) excess noise was $\approx 4.5$ mPNU with typical channel delay equivalent to $\approx 101.4$ \si{\micro\second} for the externally synchronized system. For the free-running system, we achieved an average excess noise of 4.0 mPNU. The synchronized measurements had a standard deviation of $\approx$ 10.7 mPNU, putting the free-running results well within expected variation.

Assuming an information reconciliation efficiency of $\beta=0.95$ we calculated the secret key fraction per quantum symbol in the asymptotic regime as per
\begin{equation*}
    \beta I(A:B) - \chi(B:E)\ ,
\end{equation*}
where $I(A:B)$ is the mutual information between Alice (A) and Bob (B), and $\chi(B:E)$ is the Holevo information between Bob and the eavesdropper E. For the externally synchronized measurement and the free-running one we achieved a secret key fraction of 0.0416 and 0.0466  bits per symbol, respectively. 

A further measurement set was taken on our semi-autonomous CV-QKD system \cite{Jain2022} for a demonstration with a free-running system with a channel triggered receiver instead of an externally triggered one. The measurement was performed over 10 km of SMF and yielded $\approx 2.4$ mPNU excess noise for a secret key fraction of roughly 0.0740 bits/symbol over $9\times10^{6}$ symbols with a modulation variance $\approx3.5$, assuming the same reconciliation efficiency of 0.95. In a typical run of the QKD protocol, parameter estimation would be performed on $10^{8}$ or preferably a greater number of symbols and confidence intervals would ensure $\epsilon$-security by using worst-case bounds as per \cite{Jain2021}.

We show the bit error rate (BER) achieved by the QPSK signal with respect to the excess noise in Fig.~\ref{fig:bervsepn}. It is easy to see that using the BER allows Bob to determine which frames he successfully detected and processed. This information can then be sent back to Alice allowing her to reconcile her key without the need to perform information reconciliation on those frames that would obviously not pass this step given the code rate was optimized for successfully processed frames. Out of the 2000 transmitted frames only very few failed after successful detection, showing the reliability of our method. 

\subsection{Conclusion}
In conclusion, we demonstrated a set of DSP steps that allows digital synchronization of a CV-QKD system whose local oscillator is not transmitted through the fiber but generated locally at the receiver by a free-running laser. We have experimentally shown that our proposed digital synchronization method has no penalty in terms of performance in comparison with external hardware synchronization. Our method is modulation format free allowing it to be applied to a variety of CV-QKD systems. Making use of auxiliary signals that can be generated by the same arbitrary waveform generator as the quantum states and detected by the same receiver, no additional hardware is required, thereby obsoleting sophisticated analog synchronization. As the performance of CV-QKD systems in terms of secret key rate is usually limited by the throughput of information reconciliation~\cite{Li2020} and not by the symbol rate of the quantum signal, the additional bandwidth requirement of our method is of no harm.

\section*{Acknowledgements}
The authors acknowledge financial support from Innovation Fund Denmark (CryptQ project, grant agreement no. 0175-00018A) and from European Union’s Horizon 2020 research and innovation programmes CiViQ (grant agreement no. 820466) and OPEN-QKD (grant agreement no. 857156). HMC, NJ, ULA and TG acknowledge support from the Danish National Research Foundation, Center for Macroscopic Quantum States (bigQ, DNRF142), and use of DCC computing resources \cite{DTU_DCC_resource}.


\bibliography{20210526}

\end{document}